\documentclass[11pt]{article}

\usepackage[final]{acl}
\usepackage{times}
\usepackage{latexsym}
\usepackage{amsmath}
\usepackage{hyperref}
\usepackage{graphicx}
\usepackage{algorithm}
\usepackage{algpseudocode}
\usepackage{amsmath}
\usepackage{mathrsfs}
\usepackage{amsfonts}
\usepackage{arydshln} 
\usepackage{booktabs}  
\usepackage{multirow}  
\usepackage{enumitem}  

\usepackage{times}
\usepackage{latexsym}

\usepackage[T1]{fontenc}

\usepackage[utf8]{inputenc}

\usepackage{microtype}

\usepackage{inconsolata}

\usepackage{graphicx}
\usepackage{amsmath} 
\usepackage{multicol}
\usepackage{multirow}
\usepackage{booktabs}
\newcommand{\method}[0]{GRLM}

\title{Unleashing the Native Recommendation Potential: LLM-Based Generative Recommendation via Structured Term Identifiers}

\author{
    \textbf{
        Zhiyang Zhang,
        Junda She,
        Kuo Cai,
        Bo Chen,
        Shiyao Wang,
        Xinchen Luo,
    } \\
    \textbf{
        Qiang Luo\thanks{Corresponding authors},
        Ruiming Tang\footnotemark[1],
        Han Li,
        Kun Gai,
        Guorui Zhou
    } \\
\textsuperscript{\rm 1}Kuaishou Inc., Beijing, China\\
{\tt \{zhangzhiyang06,luoqiang,tangruiming\}@kuaishou.com}\\
}

\begin{document}
\maketitle
\begin{abstract}
Leveraging the vast open-world knowledge and understanding capabilities of Large Language Models (LLMs) to develop general-purpose, semantically-aware recommender systems has emerged as a pivotal research direction in generative recommendation. However, existing methods face bottlenecks in constructing item identifiers. Text-based methods introduce LLMs' vast output space, leading to hallucination, while methods based on Semantic IDs (SIDs) encounter a semantic gap between SIDs and LLMs' native vocabulary, requiring costly vocabulary expansion and alignment training. To address this, this paper introduces \textbf{Term IDs (TIDs)}, defined as a set of semantically rich and standardized textual keywords, to serve as robust item identifiers. We propose \textbf{\method}\footnote{\url{https://github.com/ZY0025/GRLM}}, a novel framework centered on TIDs, employs Context-aware Term Generation to convert item's metadata into standardized TIDs and utilizes Integrative Instruction Fine-tuning to collaboratively optimize term internalization and sequential recommendation. Additionally, Elastic Identifier Grounding is designed for robust item mapping. Extensive experiments on real-world datasets demonstrate that \method\ significantly outperforms baselines across multiple scenarios, pointing a promising direction for generalizable and high-performance generative recommendation systems.
\end{abstract}

\section{Introduction}

\begin{figure*}[t]
  \centering
  \includegraphics[width=\linewidth]{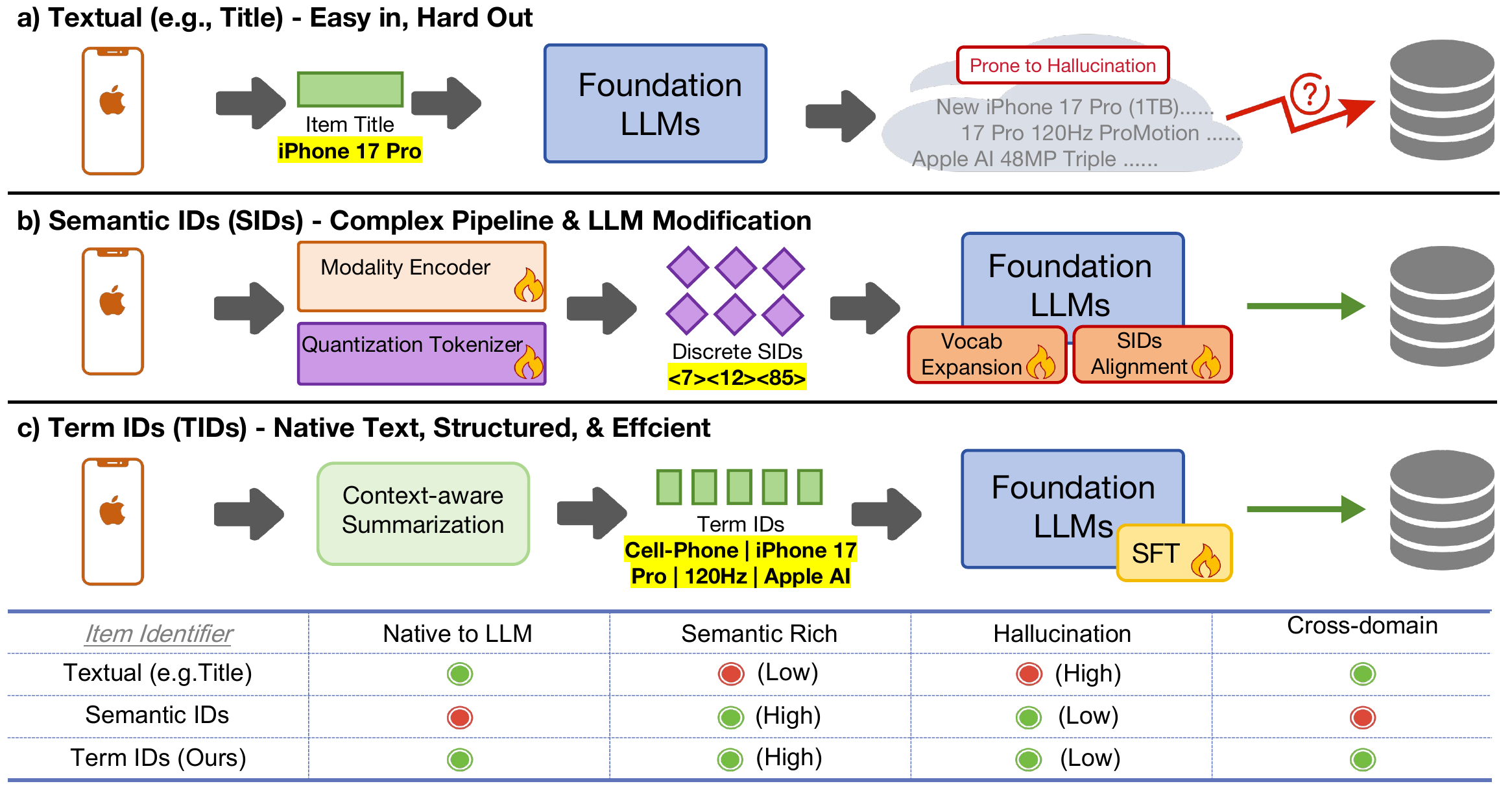}
  \caption{Comparison of Item Identifiers: Term IDs (TIDs) leverage standardized and structured text tokens to ensure precise semantic extraction and low hallucination, while maintaining native compatibility with LLMs vocabularies without the complex indexing pipelines or architectural modifications required by Semantic IDs (SIDs).}
  \label{fig:1}
\end{figure*}

Large Language Models (LLMs) have demonstrated exceptional capabilities in complex semantic understanding and generation~\cite{achiam2023gpt}, leading to paradigm shifts across various fields. In the realm of recommender systems~\cite{radford2019language, brown2020language}, a novel Generative Recommendation (GR) paradigm has emerged~\cite{wang2023generative, deldjoo2024review}, transforming the traditional multi-stage recall-and-ranking pipeline~\cite{covington2016deep, qin2022rankflow} into a generative task that auto-regressively generates item identifiers based on user historical behaviors, which has gained widespread application~\cite{zhai2024actions,deng2025onerec,han2025mtgr,huang2025towards}. 
When pretrained LLMs are employed as the backbone of GR to strengthen open-world knowledge and reasoning capabilities, a fundamental challenge lies in \textbf{Item Identifiers}: enabling LLMs to represent continuously emerging and diverse items within their parameter space, while reliably grounding generated tokens to valid real-world items during generation.

As illustrated in Figure \ref{fig:1}, existing methods primarily follow two paths. The first utilizes \textit{Textual Identifiers} (e.g., titles or brief descriptions)~\cite{chu2023leveraging, liu2025understanding, tan2024idgenrec}. While these leverage the LLM’s native vocabulary to depict items, they suffer from two major flaws.
Raw titles frequently lack sufficient discriminative information, while full descriptions are too lengthy for efficient sequence modeling, leading to semantic instability. Moreover,  due to the vast output space, LLMs are prone to hallucinations, resulting in the generation of non-existent or corrupted items. 

The second path involves \textit{Semantic IDs (SIDs)}, which quantize~\cite{lee2022autoregressive, yin2013regularized} item embeddings into discrete codes~\cite{rajput2023recommender,zheng2024adapting, yang2025sparse}. Although earlier architectures showed promise, they often function as specialized models rather than general-purpose LLMs. When attempting to integrate this paradigm with LLMs, the standard practice involves extending the LLM's native vocabulary with additional tokens specifically for SIDs~\cite{liu2025onerec_think, zhou2025openonerec}. Consequently, they encounter a semantic gap: these numerical codes are absent from the LLM’s pre-trained vocabulary, failing to tap into the model's latent world knowledge. This necessitates costly vocabulary expansion and intensive alignment training. Furthermore, since these numerical codes are inherently devoid of universal semantics, they are often domain-specific, which severely hinders the model's effectiveness in cross-domain recommendation.


To mitigate the limitations of existing item identifiers, we investigate a novel item identifier approach built upon the native vocabulary of LLMs, which requires addressing several non-trivial challenges. First, the redundancy of natural language demands effective extraction of compact yet informative terms to represent items. Second, synonym ambiguity across different lexical realizations can lead to grounding conflicts and hallucinations. Third, item identifiers should preserve semantic similarity among related items while maintaining sufficient discriminability. 

To overcome these challenges, we propose \textbf{G}enerative \textbf{R}ecommendation \textbf{L}anguage \textbf{M}odel (\textbf{\method}), a unified framework integrated with \textbf{Term IDs (TIDs)}, a structured item identifier that exploits the semantic understanding capabilities of LLMs to
encode items as keyword sequences with high information density. The \textit{Context-aware Term Generation (CTG)} process is designed to reduce grounding ambiguity and hallucination, while ensuring distinguishability among similar items.
By leveraging neighborhood-based in-context learning, the metadata of similar items is retrieved as guidance, enabling the LLMs to extract consistent terms for related items while capturing fine-grained discriminative features. 
Consequently, TIDs based on the native LLMs vocabulary provide generality, semantic richness, and robust grounding with minimal hallucination. The comparison with other item identifier approaches is presented in Figure ~\ref{fig:1}.

Based on the Term IDs, \method\ further introduces an \textit{Integrative Instruction Fine-tuning (IIFT)} training paradigm, which jointly fine-tunes LLMs on an item-to-TIDs identification task and a personalized recommendation task. This multi-task optimization enhances the LLM’s understanding of domain knowledge while strengthening its ability to capture personalized preferences, thereby enabling accurate recommendations.
Moreover, to achieve reliable item grounding during inference, a dual-level \textit{Elastic Identifier Grounding (EIG)} mechanism is introduced. By exploiting the decompositional nature of TIDs, EIG combines direct mapping with structural mapping to ground TIDs into actual items efficiently.

Our contributions can be summarized as follows:
\begin{itemize}[leftmargin=*,nosep]
  \item We propose Term IDs, a structured item identifier derived from the native LLMs vocabulary, offering generality, semantically rich, and robust grounding with minimal hallucination.
  \item We propose a LLM-based general framework \method, which integrates the item identification task and personalized recommendation task via Integrative Instruction Fine-tuning with an Elastic Identifier Grounding mechanism to ensure reliable item grounding.
  \item Extensive experiments demonstrate that \method\ achieves state-of-the-art performance across multiple benchmarks. 
  Moreover, through detailed analytical experiments, we provide evidence that our TID-based method exhibits robust scaling properties while significantly mitigating hallucination.
\end{itemize}




\section{Related Works}

\subsection{SID-based Generative Recommendation}
TIGER~\cite{rajput2023recommender} was the pioneering work to propose the generative recommendation framework. It represents items using SIDs and employs a Transformer-based~\cite{vaswani2017attention} architecture to directly generate the SID of the target item. Subsequent models, such as LETTER~\cite{wang2024learnable}, EAGER~\cite{wang2024eager}, and UNGER~\cite{xiao2025unger}, have further incorporated collaborative signals into the generation process. In specific domains, models like OneLoc~\cite{wei2025oneloc} and GNPR-SID~\cite{wang2025generative} have integrated geographic information into item embeddings. OneRec~\cite{deng2025onerec,zhou2025onerec_v1} attempts to construct an end-to-end recommendation pipeline by applying generative recommendation to large-scale industrial systems. 

Despite these advancements, traditional generative methods rely on SIDs, which lack inherent semantic interpretability and fail to fully leverage the extensive open-world knowledge and reasoning capabilities of LLMs. Although OneRec-Think~\cite{liu2025onerec_think} addresses this by proposing a unified framework to align the SID space with the natural language space, such methods typically necessitate expensive vocabulary expansion and computationally intensive alignment training.

\subsection{LLM-Based Recommendation}
Another line of research leverages the semantically rich of LLMs by representing items through textual descriptors~\cite{bao2025bi, chen2024softmax}. In this paradigm, recommendation evolves from simple pattern grounding to deep reasoning based on world knowledge. This transition not only mitigates cold-start and long-tail distribution challenges inherent in traditional collaborative filtering but also empowers systems with superior cross-domain generalization and zero-shot capabilities.

Early efforts like TallRec~\cite{bao2023tallrec} demonstrated the effectiveness of lightweight Instruction Fine-tuning, using item titles to align LLMs with recommendation tasks via specialized prompts. LLMTreeRec~\cite{zhang2025llmtreerec} organizes item attributes into a tree structure, constraining the quantization and generation process within a hierarchical framework. InteraRec~\cite{karra2024interarec} addresses text-deficient items by encoding visual information into structured textual summaries, providing a novel perspective for item tokenization. LLaRa~\cite{liao2024llara} incorporates collaborative signals by concatenating item sequence features from traditional recommenders with inherent textual attributes.

Although progress has been made, the reliability of text-based generative recommendation remains a major concern, necessitating the use of constrained decoding and fixed candidate sets. This reliance highlights hallucination, which prevents such models from achieving true generative recommendation. In contrast, our Term IDs provide a robust solution that substantially mitigates these hallucinations.

\begin{figure*}[t]
  \centering
  \includegraphics[width=\linewidth]{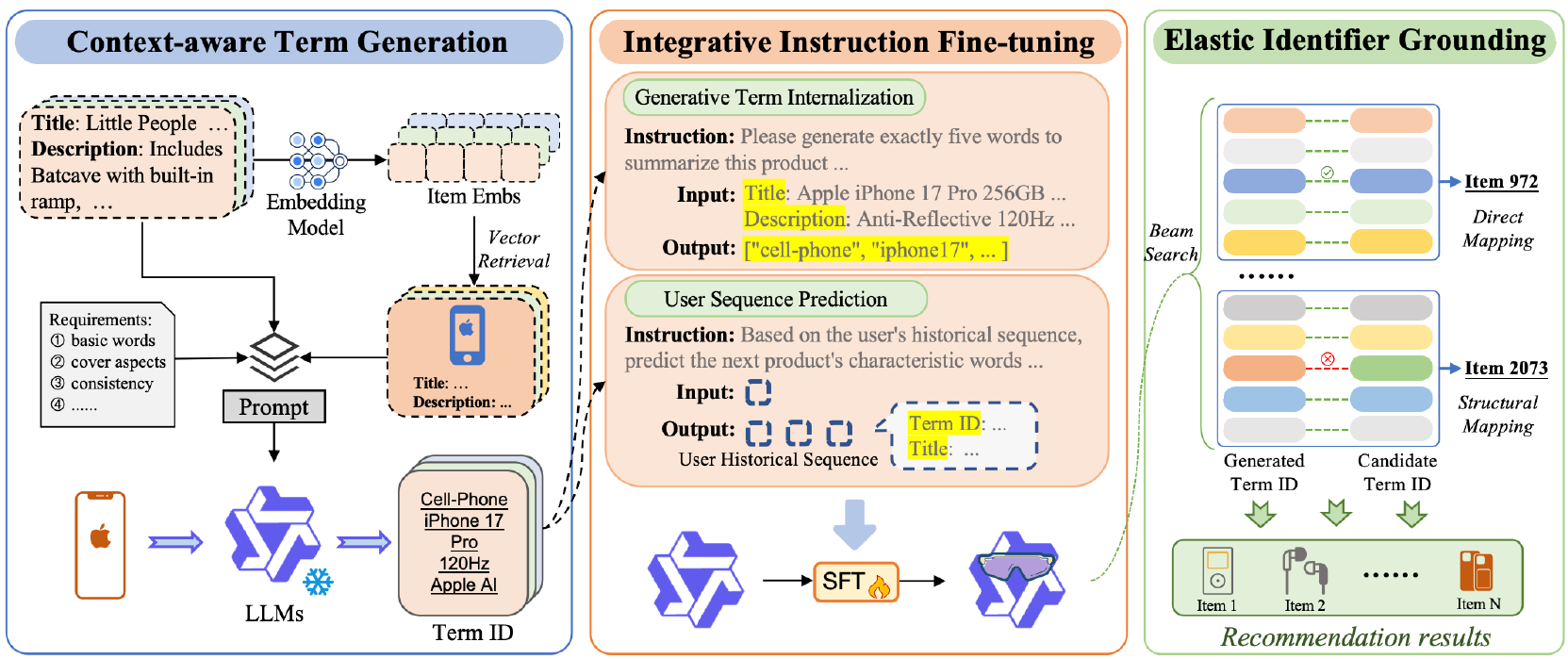}
  \caption{Overall framework of \method.}
  \label{fig:sumrec}
\end{figure*}

\section{\method}

Figure \ref{fig:sumrec} illustrates the overall framework of \method, which employs a three-stage pipeline centered on Term IDs to unleash the native recommendation capabilities of LLMs. First, we utilize \textbf{Context-aware Term Generation} to convert item's metadata into standardized TIDs; Second,  \textbf{Integrative Instruction Fine-tuning} is employed to adapt the LLMs through a multi-task learning paradigm, enabling it to internalize item semantics while simultaneously modeling user behavioral patterns. Finally, \textbf{Elastic Identifier Grounding} implements a hybrid grounding mechanism that seamlessly integrates strict direct mapping with a structural mapping, thereby ensuring robust item retrieval during the inference phase.


\subsection{Context-aware Term Generation}

Standardizing item identifiers in a generative space faces a dual challenge: maintaining term consistency across similar items to prevent identifier fragmentation, while preserving discriminative uniqueness to avoid semantic collisions between distinct products. Processing each item in isolation often fails to resolve synonymy or capture fine-grained differences. 

\begin{figure}[t]
  \centering
  \includegraphics[width=\linewidth]{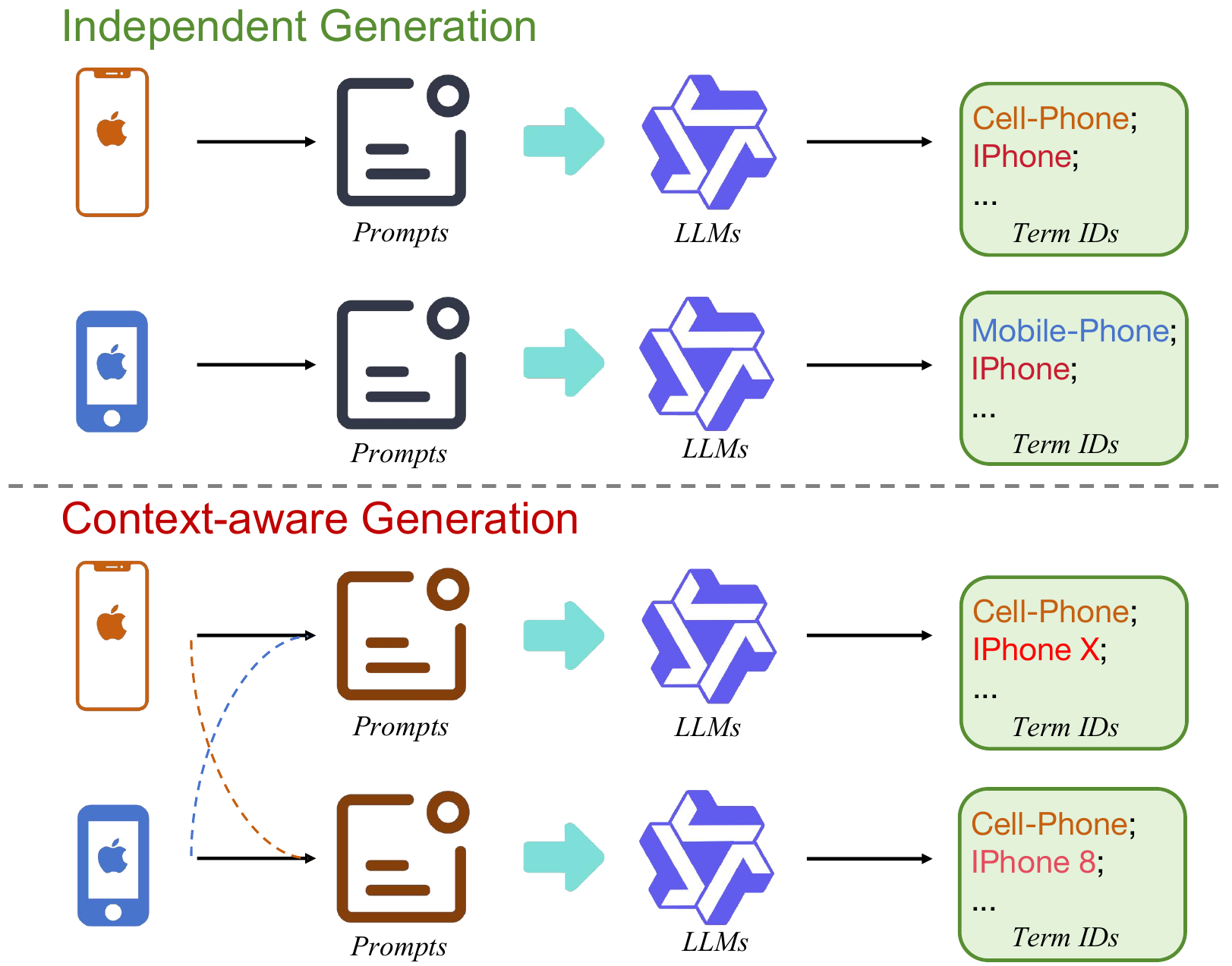}
  \caption{Context-aware Term Generation effectively ensures that Term IDs across items are globally consistent and locally discriminative.}
  \label{fig:2}
\end{figure}

As illustrated in Figure \ref{fig:2}, independent generation results in inconsistent labeling for identical features, such as "Cell-Phone" versus "Mobile-Phone," while simultaneously failing to distinguish between different models by assigning the generic "iPhone" to both. 

To address this, we propose \textbf{Context-aware Term Generation (CTG)} aims to convert item's metadata into standardized, human-readable Term IDs by incorporating similar neighborhoods as contextual guidance. For each item $i \in \mathcal{I}$, we first aggregate its metadata $m_i$ and encode it into a dense vector $\mathbf{v}_i \in \mathbb{R}^d$ using a pre-trained embedding model\footnote{https://huggingface.co/Qwen/Qwen3-Embedding-8B}.

Next, we calculate the cosine similarity between the target item $i$ and all other items $j$ in the candidate library. We retrieve the top-$k$ most similar items to form the context set $\mathcal{N}_i = \{j_1, j_2, \dots, j_k\}$. We then construct a structured prompt $\mathcal{P}$ that integrates the metadata of item $i$ and its neighbors $\{m_j\}_{j \in \mathcal{N}_i}$. The design philosophy of our prompt emphasizes a balance between globally consistency and locally discriminative: it instructs the LLMs to adopt standardized terms for shared attributes among neighbors while intentionally selecting keywords that capture the item's distinctive features, detailed prompts are provided in Appendix \ref{sec:appendix-exp-details} Figure \ref{fig:prompt}. The process is formulated as:
\begin{equation}
T_i = \text{LLM}(\mathcal{P}(m_i, \{m_j\}_{j \in \mathcal{N}_i})),
\end{equation}

By leveraging the neighborhood context, CTG enables the LLMs to recognize commonalities and maintain term consistency across similar products, effectively mitigating identifier fragmentation caused by synonyms. Simultaneously, by prompting the model to contrast an item with its neighbors, CTG enhances the capture of fine-grained discriminative information, which helps distinguish items within the same category and reduces potential semantic overlaps. 

\subsection{Integrative Instruction Fine-tuning}

Traditional method often focuses solely on sequence prediction, which may cause the model to lose the semantic grounding of the identifiers. We propose \textbf{Integrative Instruction Fine-tuning (IIFT)}, which enable the model to collaboratively optimize Generative Term Internalization (GTI) and User Behavior Sequence Prediction. 

\textbf{Generative Term Internalization} This task requires the model to directly generate the standardized $T_i$ from item's metadata $m_i$ without the neighborhood context used during the generation phase. The core motivation is to encourage the LLMs to internalize the underlying generation logic, effectively compressing its expansive natural language output space into a more focused and "distilled" semantic region. By learning to map item's metadata to standardized TIDs, the model establishes an intrinsic understanding of the identifier's structure. 

\textbf{User Behavior Sequence Prediction} As the core recommendation task, this phase models the evolution of user interests using historical interaction sequences. For a user sequence $S = \{i_1, i_2, \dots, i_n\}$, where $n$ denotes the total number of items interacted with by the user, we first represent each item $i_j$ as a joint textual sequence $x_j$ by concatenating its Term IDs and raw title:
\begin{equation}
x_j = [T_{i_j} \ ; \ m_{i_j}^{title}].
\end{equation}

The model is then trained in an autoregressive manner to predict each subsequent item in the history. Specifically, we designate the first item $x_1$ as the input context, while the subsequent textual sequence $\{x_2, \dots, x_n\}$ are concatenated to form the output. The model is trained to minimize the negative log-likelihood of the output tokens, conditioned on the initial instruction $\mathcal{I}$:
\begin{equation}
\mathcal{L} = - \sum_{k=2}^{n} \log P(x_k \ | \ \mathcal{I} (x_1,...,x_{k-1})).
\end{equation}
This formulation ensures that the model learns to reconstruct the entire behavioral trajectory by focusing its predictive capacity on the tokens following the anchor item $x_1$.

Notably, during the inference phase, the model is only required to generate the Term IDs $T_{i_{t+1}}$ for the next item, which can be mapping to the specific item, rather than the full $x_{t+1}$ containing the title, thereby ensuring high inference efficiency. 

\subsection{Elastic Identifier Grounding}
Unlike traditional generative methods that treat item identifiers as numerical indices or text strings, Term IDs represents items as structured sets of standardized semantic terms. To fully harness this architectural advantage, we design Elastic Identifier Grounding (EIG), a dual-level retrieval mechanism that mapping the generated sequence to the recommend item:

1) Direct Mapping: We first attempt an exact string-level mapping within the candidate library $\mathcal{C}$. This track ensures maximum precision when the model’s output perfectly aligns with the standardized TIDs of a specific item. 

2) Structural Mapping: If no exact mapping is found, EIG leverages the decompositional nature of TIDs to perform structural mapping. We resolve the final identifier by identifying the item $i^*$ that maximizes the structural score:
\begin{equation}
s^* = \arg\max_{i \in \mathcal{C}} \sum_{j=1}^N w_j \cdot \mathbb{I}(t_{gen}^j = t_{i}^j),
\end{equation}
where $t_{gen}^j$ is the $j$-th term in the generated sequence, and $w_j = \frac{1}{j+1}$ is a decay weight.

\section{Experiment Setup}
To comprehensively evaluate the capability and generalization of \method, we conduct experiments under the two most common scenarios in recommendation systems: in-domain and cross-domain.

\textbf{Baseline}. 
To verify the effectiveness of \method, we compare it against a diverse set of competitive baselines, categorized as follows:
\begin{itemize}[leftmargin=*]
    \item {Sequential Methods}: We include SASRec~\cite{kang2018self} and BERT4Rec~\cite{sun2019bert4rec}, which employ self-attention mechanisms to model user behavior sequences.
    \item {Generative Recommenders}: We compare against TIGER~\cite{rajput2023recommender} and HSTU~\cite{zhai2024actions}, representing  generative recommendation methods.
    \item {LLM-based Recommenders}: We include OneRec-Think~\cite{liu2025onerec_think} and IDGenRec~\cite{tan2024idgenrec}, which are representative methods that leverage LLMs for recommendation tasks.
    \item {Recommenders for Cross-Domain}: For cross-domain scenarios, we additionally include TriCDR~\cite{ma2024triple}, LLM4CDSR~\cite{liu2025bridge}, and GenCDR~\cite{hu2025ids}, which are specifically designed to transfer knowledge across domains.
\end{itemize}

\textbf{Dataset}. For the in-domain scenario, we select three real-world recommendation datasets from the popular Amazon product review dataset\footnote{https://jmcauley.ucsd.edu/data/amazon/}: \textit{Beauty}, \textit{Sports}, and \textit{Toys}. 
Following the settings in ~\cite{rajput2023recommender,liu2025onerec_think} for data pre-processing, we perform 5-core filtering and use leave-one-out strategy to split datasets. 


For the cross-domain scenario, we conduct experiments on two dataset pairs: \textit{Sports-Clothing} (Leisure) and \textit{Phones-Electronics} (Technology). 
First, each individual dataset is processed following the same filtering strategy as the in-domain scenario. 
Then, adhering to the protocols in~\cite{hu2025ids}, we construct a unified cross-domain sequence for each overlapping user by merging interactions from both datasets and sorting them chronologically. 
During the evaluation phase, we predict the last interaction for each dataset in the pair separately. 
This setup allows the model to capture dynamic preference transfer across domains within a single coherent context.
Table \ref{table:dataset} shows the dataset statistics.

\textbf{Implementation Details}. 
In our experiment, the length of the Term IDs for all items is set to 5. For \method's backbone, we select the newest Qwen3-4B-2507\footnote{https://huggingface.co/Qwen/Qwen3-4B-Instruct-2507} version as our term generation model and recommendation model, with model parameters frozen during CTG and fully fine-tuned during IIFT. Beam search strategy is used during the model's inference process. Considering the varying lengths of Term IDs for different items, the model's generation max-length is set to 30 to accommodate all Term IDs. Top-\textit{k} Recall and NDCG with K=5 and 10 are used as metrics, following ~\cite{rajput2023recommender,liu2025onerec_think}. Training details are in Appendix \ref{sec:appendix-exp-details}.

\begin{table}[t]
\centering
\caption{Statistics of the datasets used in our experiments.
}
\resizebox{1.0\linewidth}{!}{
\begin{tabular}{clccccccccccc}
\hline
\multirow{2}{*}{Dataset} & \multicolumn{3}{c}{In-domain} &  \multicolumn{4}{c}{Cross-domain}  \\ \cmidrule(lr){2-4} \cmidrule(lr){5-8} 
& Beauty & Sports & Toys & Sports & Clothing & Phones & Electronics\\
\hline
\#User & 22,363 & 35,598 & 19,412 & 35,598 & 39,387 & 27,879 & 192,403\\
\#Item & 12,101 & 18,357 & 11,924 & 18,357 & 23,033 & 10,429 & 63,001\\
\hline
\end{tabular}
}
\label{table:dataset}
\end{table}



\section{Experiment Result}

To comprehensively evaluate the effectiveness and robustness of \method, our experiments are designed to address the following research questions:
\begin{itemize}[leftmargin=*]
    \item \textbf{RQ1:} How does \method\ perform compared to state-of-the-art baselines in both in-domain and cross-domain recommendation scenarios?
    \item \textbf{RQ2:} What is the contribution of core components, such as Context-aware Term Generation and Integrative Instruction Fine-tuning, to the overall performance?
    \item \textbf{RQ3:} Does \method\ exhibit positive scaling properties consistent with the scaling laws of Large Language Models?
    \item \textbf{RQ4:} Can the proposed Term IDs effectively mitigate the hallucination inherent in text-based generative recommendation?
\end{itemize}

\subsection{Overall Performance (RQ1)}

\begin{table*}[t]
\centering
\caption{Overall performance comparison on different datasets (In-domain). The best and second-best results are highlighted in \textbf{bold} font and \underline{underlined}.}

\resizebox{1.0\linewidth}{!}{
\begin{tabular}{clcccccccccc}
\hline
\multirow{2}{*}{Dataset} & \multicolumn{1}{l}{\multirow{2}{*}{Metric}} & \multicolumn{2}{c}{Sequential methods} &  \multicolumn{2}{c}{Generative} &  \multicolumn{2}{c}{LLM-Based} & \multirow{2}{*}{\textbf{\method}} & \multirow{2}{*}{Improvement} \\ \cmidrule(lr){3-4} \cmidrule(lr){5-6} \cmidrule(lr){7-8}
\multicolumn{2}{c}{} & SASRec & BERT4Rec & TIGER & HSTU & IDGenRec & OneRec-Think   &  \\ \hline
\multirow{4}{*}{Beauty} 
&Recall@5  &0.0402&0.0232 &0.0405&0.0424 & 0.0484 & \underline{0.0563} & \textbf{0.0607}&7.82\%\\
&Recall@10 &0.0607&0.0396 &0.0623&0.0652 & 0.0693 & \underline{0.0791} & \textbf{0.0846}&6.95\%\\
&NDCG@5   &0.0254&0.0146 &0.0267&0.0280 & 0.0337 & \underline{0.0398} & \textbf{0.0430}&8.04\%\\
&NDCG@10  &0.0320&0.0199 &0.0337&0.0353 & 0.0404 & \underline{0.0471} & \textbf{0.0506}&7.43\%\\
\hline
\multirow{4}{*}{Sports}  
&Recall@5  &0.0199&0.0102 &0.0215&0.0268 & 0.0270 & \underline{0.0288} & \textbf{0.0375}&30.21\%\\
&Recall@10 &0.0301&0.0175 &0.0347&0.0343 & 0.0388 & \underline{0.0412} & \textbf{0.0539}&30.83\%\\
&NDCG@5   &0.0106&0.0065 &0.0137&0.0173 & 0.0185 & \underline{0.0199} & \textbf{0.0260}&30.65\%\\
&NDCG@10  &0.0141&0.0088 &0.0179&0.0226 & 0.0223 & \underline{0.0239} & \textbf{0.0313}&30.96\%\\
\hline
\multirow{4}{*}{Toys}  
&Recall@5  &0.0448&0.0215 &0.0337&0.0366 & \underline{0.0595} & 0.0579 & \textbf{0.0684}&14.96\%\\
&Recall@10 &0.0626&0.0332 &0.0547&0.0566 & \underline{0.0800} & 0.0797 & \textbf{0.0942}&17.75\%\\
&NDCG@5   &0.0300&0.0131 &0.0209&0.0245 & \underline{0.0432} & 0.0412 & \textbf{0.0477}&10.42\%\\
&NDCG@10  &0.0358&0.0168 &0.0276&0.0309 & \underline{0.0498} & 0.0482 & \textbf{0.0561}&12.65\%\\
\hline
\end{tabular}
}
\label{table:indomain}
\end{table*}

\begin{table*}[t]
\centering
\caption{Overall performance comparison on different datasets (Cross-domain). The best and second-best results are highlighted in \textbf{bold} font and \underline{underlined}.}
\resizebox{1.0\linewidth}{!}{
\begin{tabular}{cclccccccccccc}
\hline
\multirow{2}{*}{Scenario} & \multirow{2}{*}{Dataset} & \multicolumn{1}{l}{\multirow{2}{*}{Metric}} & \multicolumn{2}{c}{Sequential methods} & \multicolumn{2}{c}{Generative} & \multicolumn{3}{c}{Cross-Domain} & \multirow{2}{*}{\textbf{\method}} & \multirow{2}{*}{Improvement} \\ \cmidrule(lr){4-5} \cmidrule(lr){6-7} \cmidrule(lr){8-10}
\multicolumn{3}{c}{} & SASRec & BERT4Rec  & TIGER & HSTU & TriCDR & LLM4CDSR & GenCDR &  \\ \hline
\multirow{8}{*}{\rotatebox{90}{Leisure}} & \multirow{4}{*}{Sports} 
& Recall@5  & 0.0188 & 0.0197  & 0.0267 & 0.0254 & 0.0266 & 0.0263 & \underline{0.0274} & \textbf{0.0480}  & 75.18\%  \\
& & Recall@10 & 0.0325 & 0.0334  & 0.0397 & 0.0381 & 0.0396 & 0.0398 & \underline{0.0403} & \textbf{0.0645} & 60.05\%  \\
& & NDCG@5    & 0.0121 & 0.0126  & 0.0244 & 0.0241 & 0.0255 & 0.0257 & \underline{0.0261} & \textbf{0.0353} & 35.25\%  \\
& & NDCG@10   & 0.0169 & 0.0173  & \underline{0.0287} & 0.0277 & 0.0259 & 0.0260  & 0.0262 & \textbf{0.0406} & 41.46\%  \\
\cline{2-13}
& \multirow{4}{*}{Clothing}  
& Recall@5  & 0.0128 & 0.0132  & 0.0173 & 0.0175 & 0.0174 & 0.0176 & \underline{0.0181} & \textbf{0.0288} & 59.12\%  \\
& & Recall@10 & 0.0219 & 0.0227  & 0.0241 & 0.0253 & 0.0258 & 0.0261 & \underline{0.0265} & \textbf{0.0436} & 64.53\% \\
& & NDCG@5    & 0.0078 & 0.0081  & 0.0125 & 0.0132 & 0.0161 & 0.0163 & \underline{0.0167} & \textbf{0.0190}  & 13.77\%   \\
& & NDCG@10   & 0.0105 & 0.0108  & 0.0167 & 0.0174 & 0.0194 & 0.0196 & \underline{0.0203} & \textbf{0.0238} & 17.24\%   \\
\hline
\multirow{8}{*}{\rotatebox{90}{Technology}} & \multirow{4}{*}{Phones}  
& Recall@5  & 0.0331 & 0.0345 & 0.0423 & 0.0415 & 0.0434 & 0.0431 & \underline{0.0436} & \textbf{0.0928} & 112.84\% \\
& & Recall@10 & 0.0524 & 0.0537 & 0.0613 & 0.0615  & 0.0593 & 0.0614 & \underline{0.0621} & \textbf{0.1172} & 88.73\%   \\
& & NDCG@5    & 0.0215 & 0.0224 & 0.0315 & 0.0327  & 0.0396 & 0.0401 & \underline{0.0411} & \textbf{0.0739} & 79.81\%  \\
& & NDCG@10   & 0.0278 & 0.0287 & 0.0406 & 0.0425 & 0.0505 & 0.0506 & \underline{0.0512} & \textbf{0.0818} & 59.77\%  \\
\cline{2-13}
& \multirow{4}{*}{Electronics}  
& Recall@5  & 0.0179 & 0.0186  & 0.0228 & 0.0232 & 0.0238 & 0.0237 & \underline{0.0241} & \textbf{0.0377} & 56.43\%   \\
& & Recall@10 & 0.0276 & 0.0285  & 0.0322 & 0.0328 & 0.0339 & 0.0338 & \underline{0.0342} & \textbf{0.0529} & 54.68\%   \\
& & NDCG@5    & 0.0118 & 0.0122  & 0.0214 & 0.0226 & 0.0231 & 0.0230  & \underline{0.0235} & \textbf{0.0268} & 14.04\%  \\
& & NDCG@10   & 0.0149 & 0.0154  & 0.0269 & 0.0271 & \underline{0.0280}  & 0.0279 & 0.0283 & \textbf{0.0317} & 12.01\%   \\
\hline
\end{tabular}
}
\label{table:crossdomain}
\end{table*}

The results in both in-domain and cross-domain scenarios are shown in Table \ref{table:indomain} and Table \ref{table:crossdomain}. In the in-domain scenario, \method\ based on Term IDs achieved optimal performance across all evaluation metrics on the three datasets, significantly surpassing all baseline methods. Specifically, on the Beauty, Sports, and Toys datasets, \method\ achieved relative improvements of 7.8\%, 30.2\%, and 14.9\% in Recall@5 compared to the strongest baseline model. This strongly demonstrates the effectiveness of our proposed Term IDs-based recommendation paradigm in unleashing the native recommendation capabilities of LLMs.

A key observation in Table \ref{table:crossdomain} is \method's remarkable performance in cross-domain scenarios, with Recall@K improvements exceeding 50\% on average. Notably, \method\ achieves this without any specific architecture or auxiliary alignment modules for cross-domain required by methods like TriCDR or GenCDR. This stems from the textual nature of TIDs: by mapping items to a universal semantic space rather than domain-constrained IDs, \method\ leverages the LLM's pre-trained world knowledge to facilitate seamless knowledge transfer. The shared vocabulary of natural language act as a "semantic bridge," allowing the model to recognize functional similarities (e.g., from "Phones" to "Electronics").


\subsection{Ablation Study (RQ2)}

\begin{table}[t]
\centering
\caption{Ablation study analyzing the contribution of CTG and GTI on in-domain datasets.
}
\resizebox{1.0\linewidth}{!}{
\begin{tabular}{clccccccccccc}
\hline
{Dataset} & Metric & w/o CTG & w/o GTI & \textbf{\method}  \\ \hline
\multirow{4}{*}{Beauty} 
& Recall@5 & 0.0576 & 0.0564 & \textbf{0.0607} \\
& Recall@10 & 0.0810 & 0.0830 & \textbf{0.0846} \\
& NDCG@5 & 0.0402 & 0.0398 & \textbf{0.0430} \\
& NDCG@10 & 0.0477 & 0.0483 & \textbf{0.0506} \\
\hline
\multirow{4}{*}{Sports}  
& Recall@5 & 0.0346 & 0.0361 & \textbf{0.0375} \\
& Recall@10 & 0.0495 & 0.0531 & \textbf{0.0539} \\
& NDCG@5 & 0.0233 & 0.0250 & \textbf{0.0260} \\
& NDCG@10 & 0.0281 & 0.0305 & \textbf{0.0313} \\
\hline
\multirow{4}{*}{Toys}  
& Recall@5 & 0.0637 & 0.0653 & \textbf{0.0684} \\
& Recall@10 & 0.0857 & 0.0889 & \textbf{0.0942} \\
& NDCG@5 & 0.0458 & 0.0463 & \textbf{0.0477} \\
& NDCG@10 & 0.0529 & 0.0539 & \textbf{0.0561} \\
\hline
\end{tabular}
}
\label{table:ablation}
\end{table}

\begin{figure*}[th]
  \includegraphics[width=\linewidth]{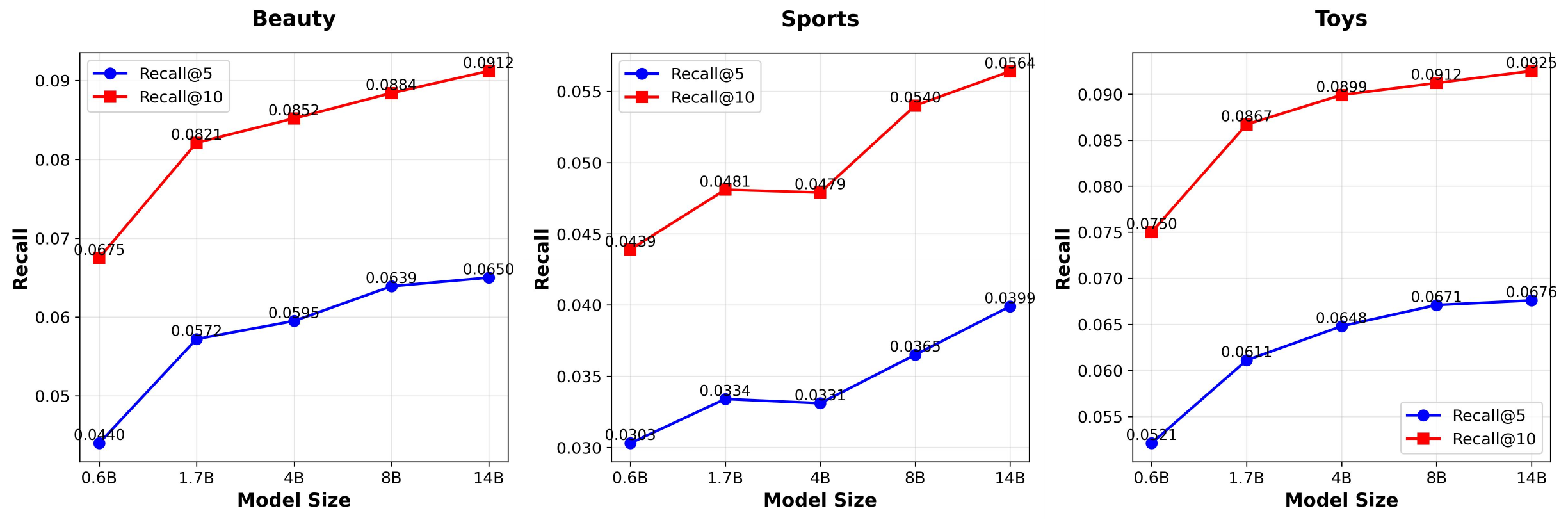} \hfill
  \caption {Performance scaling of \method\ with respect to different backbone model sizes (from 0.6B to 14B) on three in-domain datasets.} 
  \label{fig:scaling}
\end{figure*}

We conduct an ablation study on the in-domain scenario, comparing two configurations:  
1) \textbf{w/o CTG}: generated Term IDs without using similar items as context.
2) \textbf{w/o GTI}: fine-tuning LLMs without the GTI task.
The results in Table \ref{table:ablation} validate the efficacy of each component in \method. Removing CTG (w/o CTG) leads to a noticeable performance decline, confirming that leveraging neighborhood information as a contextual reference is crucial for generating locally discriminative and globally consistent TIDs. Without such context, the generation process tends to produce generic or fragmented terms; CTG refines this "semantic tokenization," providing the recommender with more precise and high-quality identifier.

The performance drop in the w/o GTI variant highlights the importance of the Generative Term Internalization task. This task  training it to map complex metadata into a constrained TIDs space. The synergy between term internalization and sequential recommendation ensures that the model not only learns accurate item representations but also maintains reliable transitions within the semantic space.


\subsection{Scaling Law (RQ3)}

We further investigated the impact of model parameter size on the performance of \method\ to verify its Scaling Law. To ensure consistency, we continue to use the Qwen3-4B-2507 version as our term generation model, but employ the more diverse Qwen3-2504\footnote{https://huggingface.co/Qwen/Qwen3-4B} version as the backbone for fine-tuning, with model sizes covering five configurations: 0.6b, 1.7b, 4b, 8b, and 14b. The overall trend of the Recall on the three in-domain datasets is shown in Figure \ref{fig:scaling}. We found that recommendation performance steadily improves with the increase in the number of parameters of the foundational LLMs, consistent with the Scaling Law in the LLMs field. This is an exciting discovery, indicates that TIDs allow larger LLMs to better utilize their superior semantic reasoning and open-world knowledge, promising further gains with future model iterations.

\subsection{Hallucination of Term IDs (RQ4)}

\begin{table}[t]
\centering
\caption{Analysis of hallucination rates of Term IDs generated by \method\ across in-domain and cross-domain datasets. We use valid rate and direct hit rate as metrics.}
\resizebox{1.0\linewidth}{!}{
\begin{tabular}{clccccccccccc}
\hline
\multirow{2}{*}{Dataset} & \multicolumn{3}{c}{In-domain} &  \multicolumn{2}{c}{Out-domain}  \\ \cmidrule(lr){2-4} \cmidrule(lr){5-6} 
& Beauty & Sports & Toys & Leisure  & Technology \\
\hline
\#VR@10 & 0.996 & 0.989 & 0.995 & 0.998 & 0.998 \\
\#DHR@10 & 0.997 & 0.999 & 0.998 & 0.999 & 1 \\
\hline
\end{tabular}
}
\label{table:Hallucination}
\end{table}

Although TIDs reside in the LLM's expansive generative space, the \method\ framework effectively constrains the output to a specific semantic subspace. To quantify this, we evaluate two metrics: Valid Rate (VR@K), the proportion of generated identifiers belonging to the candidate library, and Direct Hit Rate (DHR@K), the proportion of successful retrievals handled by the Direct Mapping track within EIG (i.e., exact direct mapping before invoking structural mapping).

As shown in Table \ref{table:Hallucination}, the model trained with \method\ exhibits exceptional stability, with both VR@10 and DHR@10 consistently exceeding 99\% across all datasets. These results indicate that \method\ successfully internalizes the structural and semantic constraints of TIDs. By learning to navigate this predefined subspace, the model significantly mitigates the hallucination bottleneck inherent in traditional text-based generative methods, achieving high grounding precision without sacrificing generative flexibility.

\section{Conclusion}
In this paper, we introduced \method, a LLM-based generative recommendation framework that utilizes Term IDs to unleash the native recommendation capabilities of LLMs. By representing items as structured TIDs within the LLM's parameter space, \method\ eliminates the need for vocabulary expansion and complex cross-modal alignment. 
Through Context-aware Term Generation, Integrative Instruction Fine-tuning, and Elastic Identifier Grounding, \method\ achieves state-of-the-art performance across both in-domain and cross-domain scenarios. Our analysis demonstrates that \method\ not only mitigates hallucination with near-perfect reliability but also exhibits strong scaling properties and cross-domain transferability. By utilizing the LLM's native vocabulary, \method\ eliminates the need for costly vocabulary expansion and alignment. Our work providing a promising and generalizable direction for high-performance generative recommendation systems.

\section{Limitations}
The following are some limitations that \method\ may face: (1) Our Context-aware Term Generation (CTG) currently utilizes a fixed external embedding model to retrieve item context. While effective, we expect to further enhance the discriminative power of Term IDs by exploring more advanced domain-specific retrieval methods in future studies. (2) Due to computational resource constraints, we primarily validated \method\ on Qwen3. Given its robust performance, we anticipate that larger base models will further unlock the framework’s potential, which we will investigate in our future research to improve universal recommendation capabilities.

\section{Ethical Statements}
Our study do not carry any ethical concerns. Specifically, Our training data are publicly available and designated for research purposes only. We inspect our dataset to ensure it does not contain any unethical content, private information and offensive topics. Moreover, the base models we used are also publicly available for research purpose.

\bibliography{custom}

\appendix
\clearpage


\section{Experiment details}
\label{sec:appendix-exp-details}

\begin{table*}[t]
\centering
\caption{10 Amazon14 dataset statistics. }
\resizebox{1.0\linewidth}{!}{
\begin{tabular}{clccccccccccc}
\hline
\textbf{Dataset} 
    & \textbf{Baby}
    & \textbf{Beauty}
    & \textbf{Cell}
    & \textbf{Grocery}
    & \textbf{Health}
    & \textbf{Home}
    & \textbf{Pet}
    & \textbf{Sports}
    & \textbf{Tools}
    & \textbf{Toys}  &  \\ \hline
\textbf{Items}
&7,050  & 12,101  & 10429 & 8,713 & 18,534 & 28,237 & 8,510 & 18,357 & 10,217 & 11,924 \\

\textbf{Users}
&19,445  & 22,363  & 27,879 & 14,681 & 38,609 & 66,519 & 19,856 & 35,598 & 16,638 & 19,412 \\

\textbf{Interactions}
&160,792  & 198,502  & 194,439 & 151,254 & 346,355 & 551,682 & 157,836 & 296,337 & 134,476 & 167,597 \\

\hline
\end{tabular}
}
\label{table:7}
\end{table*}

\begin{table*}[t]
\centering
\caption{10 Amazon14 dataset performance.}
\resizebox{1.0\linewidth}{!}{
\begin{tabular}{lccccccccccc}
\hline
\textbf{Dataset} 
    & \textbf{Baby}
    & \textbf{Beauty}
    & \textbf{Cell}
    & \textbf{Grocery}
    & \textbf{Health}
    & \textbf{Home}
    & \textbf{Pet}
    & \textbf{Sports}
    & \textbf{Tools}
    & \textbf{Toys}  
    & \textbf{Avg.}  \\ \hline
\textbf{Recall@5}
&0.0317 & 0.0601 & 0.0656 & 0.0704 & 0.0527 & 0.0260 & 0.0554 & 0.0372 & 0.0408 & 0.0682 & 0.0463 \\

\textbf{Recall@10}
&0.0439 & 0.0854 & 0.0951 & 0.1018 & 0.0753 & 0.0376 & 0.0820 & 0.0530 & 0.0599 & 0.0940 & 0.0664 \\

\textbf{NDCG@5}
&0.0220 & 0.0424 & 0.0444 & 0.0472 & 0.0374 & 0.0182 & 0.0381 & 0.0257 & 0.0291 & 0.0481 & 0.0322 \\

\textbf{NDCG@10}
&0.0260 & 0.0505 & 0.0539 & 0.0573 & 0.0446 & 0.0220 & 0.0467 & 0.0308 & 0.0352 & 0.0565 & 0.0387 \\
\hline
\end{tabular}
}
\label{table:8}
\end{table*}

\begin{figure*}[t]
  \centering
  \includegraphics[width=\linewidth]{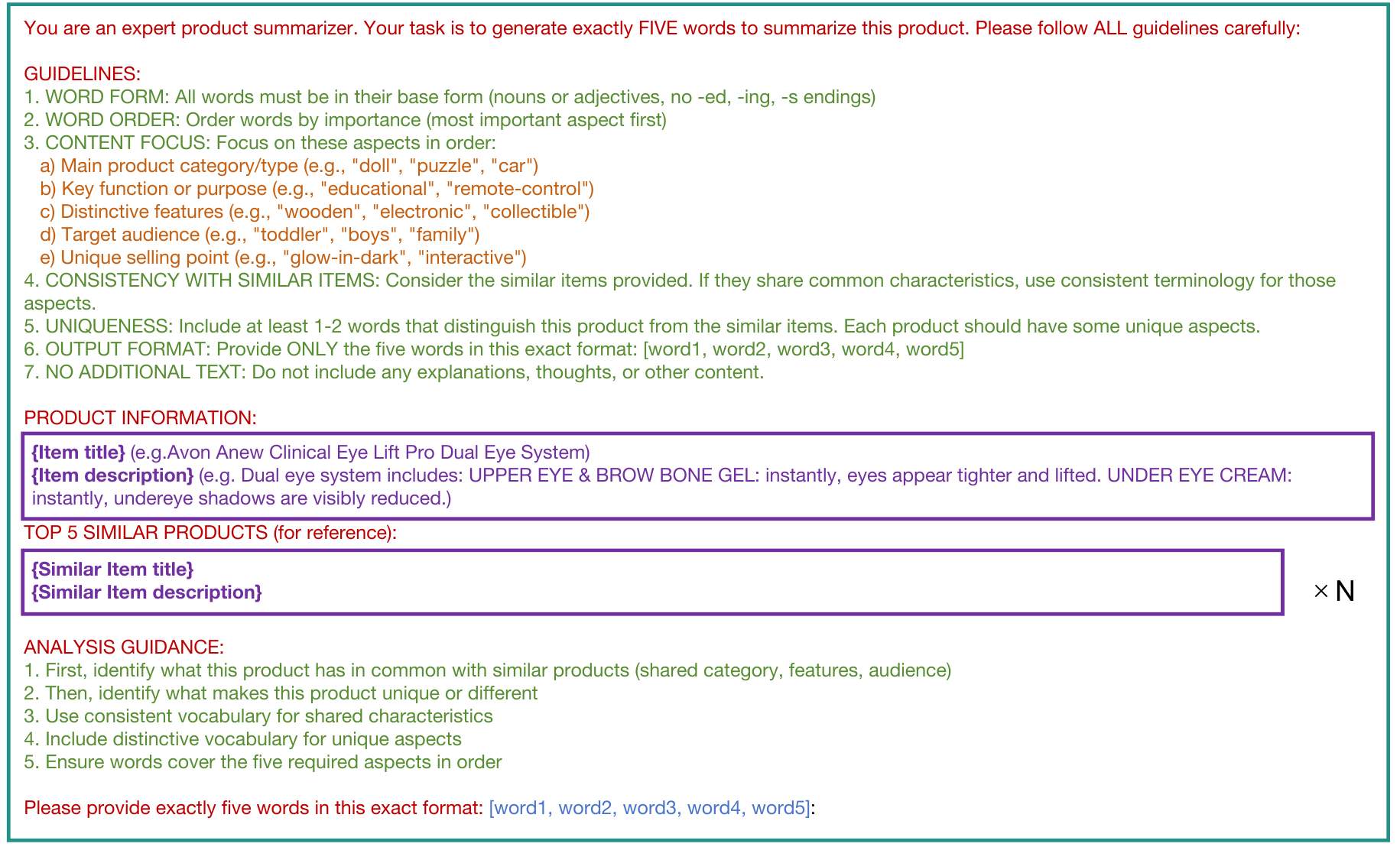}
  \caption{Prompt for Context-aware Term Generation.}
  \label{fig:prompt}
\end{figure*}

\textbf{Prompt}
To extract and synthesize key data, a structured prompt was developed to guide the LLM in summarizing specific items, as shown in Figure \ref{fig:prompt}. The prompt employs a schema-based approach, ensuring that information regarding item is condensed into a standardized format while maintaining factual density and categorical clarity.

\textbf{Training Details}
We conduct full-parameter supervised fine-tuning (SFT) on the Qwen3 models. All model variants are trained with a next-token prediction objective. We employ a cosine annealing learning rate schedule with an initial learning rate of $1 \times 10^{-4}$ and a global batch size of 128 over 3 epochs. Specifically, for the experiments in Section 3, the learning rates for the 0.6B, 1.7B, 4B, 8B, and 14B models are adjusted to $2 \times 10^{-4}$, $2 \times 10^{-4}$, $1 \times 10^{-4}$, $7 \times 10^{-5}$, and $5 \times 10^{-5}$, respectively.

\section{Length of Term IDs}

\begin{table}[t]
\caption{Length of Term IDs}
\resizebox{1.0\linewidth}{!}{
\begin{tabular}{clccccccccccc}
\hline
{Dataset} & Metric & 5 IDs & 7 IDs & 10 IDs  \\ \hline
\multirow{4}{*}{Beauty} 
& Recall@5 & 0.0607 & 0.0595 & 0.0571 \\
& Recall@10 & 0.0846 & 0.0853 & 0.0752 \\
& NDCG@5 & 0.0430 & 0.0429 & 0.0387 \\
& NDCG@10 & 0.0506 & 0.0504 & 0.0427 \\
\hline

\multirow{4}{*}{Sports} 
& Recall@5 & 0.0375 & 0.0384 & 0.0354 \\
& Recall@10 & 0.0539 & 0.0544 & 0.0525 \\
& NDCG@5 & 0.0260 & 0.0275 & 0.0249 \\
& NDCG@10 & 0.0313 & 0.0333 & 0.0297 \\
\hline

\multirow{4}{*}{Beauty} 
& Recall@5 & 0.0684 & 0.0682 & 0.0629 \\
& Recall@10 & 0.0942 & 0.0951 & 0.0875 \\
& NDCG@5 & 0.0477 & 0.0466 & 0.0428 \\
& NDCG@10 & 0.0561 & 0.0557 & 0.0507 \\
\hline
\end{tabular}
}
\label{table:termlength}
\end{table}

While the sequence-based nature of natural language suggests that a sufficiently large Term IDs space is necessary to avoid potential item collisions, it remains unclear whether further increasing the sequence length yields diminishing returns. To explore the impact of extended identifiers, we conducted experiments using 7 and 10 Term IDs per item. The results across three in-domain datasets, as summarized in Table \ref{table:termlength}, indicate that performance with 7 Term IDs remains comparable to the baseline (5 Term IDs), while increasing the count to 10 even leads to a slight decline in effectiveness. We attribute this marginal decrease to the possibility that over-extending the term generation process introduces redundant noise, thereby obscuring the item's most salient features. Given that longer sequences also increase the inference latency, we empirically select a configuration of 5 Term IDs per item to maintain an optimal trade-off between recommendation quality and computational efficiency.

\section{Case Study}

\begin{figure*}[t]
  \centering
  \includegraphics[width=\linewidth]{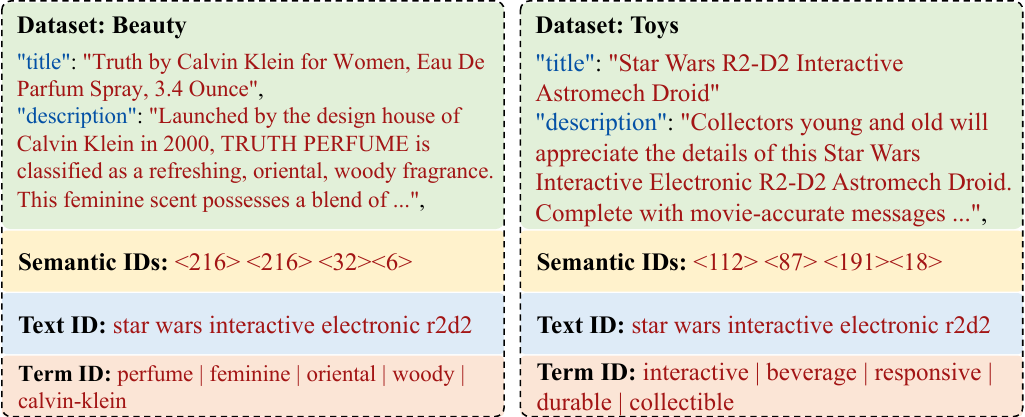}
  \caption{Case Study.}
  \label{fig:case}
\end{figure*}

As shown in Figure \ref{fig:case}, case studies across different datasets demonstrate the core advantages of our Term IDs over Semantic IDs and the Text ID generated by IDGenRec~\cite{tan2024idgenrec}. Compared to existing project identifiers, item's Term IDs effectively capture and summarize its core semantic features (such as main category, function, characteristics, etc.) in natural language. This semantic representation, built directly from the LLM's native vocabulary, not only makes the recommendation process more transparent and understandable to humans but, more importantly, helps the model better utilize the world knowledge and semantic understanding capabilities acquired by the LLMs during pre-training. This enhances the model's ability to capture user sequential patterns and generalize. This semantic consistency is the fundamental reason why Term IDs exhibit strong generalization capabilities in both in-domain and cross-domain recommendation scenarios.

\section{Robustness and Data Scalability}
To further investigate the practical potential of \method, we evaluate its performance under two challenging scenarios: large-scale joint training and semantic space compression. These experiments aim to verify whether the TID-based paradigm remains effective as the item catalog scales toward larger dimensions.

We scale the data volume by merging 10 diverse Amazon sub-datasets for joint training and evaluation (statistics and per-dataset results are detailed in Table \ref{table:7} \& \ref{table:8}). As shown in the "Avg." column of Table 8, \method\ achieves a stable Recall@10 of 0.0664 and NDCG@10 of 0.0387.

Compared to training on individual datasets (Table \ref{table:indomain}), \method\ demonstrates remarkable stability rather than the performance degradation often caused by increased item collisions. This robustness stems from the semantic grounding of TIDs: unlike numerical Semantic IDs, which may suffer from "id-space crowding" as data grows, TIDs utilize the shared natural language vocabulary across domains. For instance, common terms (e.g., "Portable," "Ergonomic") act as semantic bridges, allowing the LLMs to benefit from positive transfer across different categories, effectively increasing the training signal density for each term.

In real-world applications, the number of unique terms extracted via \method\ may grow significantly. To ensure a bounded output vocabulary, we propose a Semantic Compression strategy. We first project all extracted terms into a latent space using an embedding model and apply K-means clustering to identify $K$ cluster centroids, which we define as "Core Terms". Each original term in a TIDs is then mapped to its nearest Core Term, generating a compact set of Compressed Term IDs that maintains semantic representativeness within a fixed vocabulary size.

\begin{table}[t]
\caption{Performance of \method\ under different TIDs vocabulary sizes (compressed from 54,255 raw terms) on the merged 10-dataset.}
\resizebox{1.0\linewidth}{!}{
\begin{tabular}{clccccccccccc}
\hline
{Dataset} & Metric & K=3000 & K=5000 & K=8000 & Raw(54,255)  \\ \hline
\multirow{4}{*}{Avg.} 
& Recall@5 & 0.0457 & 0.0458 & 0.0460 & 0.0463 \\
& Recall@10 & 0.0655 & 0.0655 & 0.0657 & 0.0664 \\
& NDCG@5 & 0.0317 & 0.0319 & 0.0320 & 0.0322 \\
& NDCG@10 & 0.0377 & 0.0378 & 0.0379 & 0.0387 \\
\hline
\end{tabular}
}
\label{table:termcompression_main}
\end{table}

As illustrated in Table \ref{table:termcompression_main}, even when the term vocabulary is compressed to $K=3,000$ (approx. $1/18$ of the original size), the performance decline remains negligible ($< 1\%$). This critical finding suggests that \method\ does not rely on memorizing specific, fine-grained strings. Instead, it captures the high-level semantic abstractions of items. The ability to maintain high performance with a compact, fixed-size vocabulary highlights the scalability of the TIDs paradigm for massive-scale item catalogs.

\end{document}